\documentclass[twocolumn,aps,showpacs]{revtex4}
\usepackage[utf8]{inputenc}
\usepackage{color}
\usepackage{amsmath}
\usepackage{amssymb}
\usepackage{graphicx}

\makeatletter
\@ifundefined{textcolor}{}
{%
 \definecolor{BLACK}{gray}{0}
 \definecolor{WHITE}{gray}{1}
 \definecolor{RED}{rgb}{1,0,0}
 \definecolor{GREEN}{rgb}{0,1,0}
 \definecolor{BLUE}{rgb}{0,0,1}
 \definecolor{CYAN}{cmyk}{1,0,0,0}
 \definecolor{MAGENTA}{cmyk}{0,1,0,0}
 \definecolor{YELLOW}{cmyk}{0,0,1,0}
}

\usepackage{dcolumn}
\usepackage{bm}
\usepackage{wrapfig}
\usepackage{subfigure}
\usepackage{ucs}
\usepackage{lineno}
\usepackage{epsfig}
\usepackage{psfrag}\usepackage{epstopdf}
\DeclareGraphicsExtensions{.pdf,.eps,.png,.jpg,.mps}

\makeatother

\begin{document}
\title{Thermal entanglement and quantum coherence of a single electron in a double quantum dot with Rashba interaction}
\author{Merynilda Ferreira$^{1}$, Onofre Rojas$^{1}$, Moises Rojas$^{1}$}
\affiliation{$^{1}$Departamento de Física, Instituto de Ciências Naturais, Universidade Federal de Lavras, 37200-900,
Lavras-MG, Brazil}
\begin{abstract}
In this work, we study the thermal quantum coherence and fidelity in a semiconductor double quantum dot.  The device consists of a single 
electron in a double quantum dot with  Rashba spin-orbit coupling 
in the presence of an external magnetic field.  In our scenario, the 
thermal entanglement of the single electron is driven by the 
charge and spin qubits, the latter controlled by Rashba coupling.  
Analytical expressions are obtained for thermal concurrence and correlated coherence using  the density matrix formalism.
The main goal of this work is to provide a good understanding of the effects of temperature and several parameters in 
quantum coherence.  In addition,  our findings show that we can use the Rashba coupling to tune in the thermal entanglement, quantum coherence, as well as, the thermal fidelity behavior of the system. Moreover, we focus on the role played by thermal entanglement and correlated coherence responsible for quantum correlations. 
We observe that the correlated coherence is more robust than the thermal entanglement in all cases,  
so quantum algorithms based only on correlated coherence may be stronger than those based on entanglement. 
\end{abstract}
\maketitle

\section{introduction}

%%%%%%%%%%%%%%%%%%%%%%%%%%%%%%%
The quantum resources theories have been identified as an important field of research over the past few years \cite{chi,stre1}. In particular, 
quantum coherence and quantum entanglement represent  two fundamental features of non-classical systems that can 
each be characterized within an operational resource theory for quantum technological applications in the context of quantum 
information process \cite{benn-1,Ben2,lamico} and emerging fields such as 
quantum metrology \cite{fro,gio}, quantum thermodynamics \cite{brandao,lan} and quantum biology \cite{lambert}.
Furthermore, over the past decade, the manipulation and generation of quantum correlations, has been widely investigated 
on various quantum systems  such as Heisenberg models \cite{arne,kam,ro,ro-1}, trapped ions \cite{tur}, cavity quantum 
electrodynamics \cite{rai,davi} and so on. 

One of the most promising physical systems for implementing quantum technologies, particularly quantum computing, 
is solid-state quantum dots (QDs) \cite{peta,shin}. There are proposals for QDs devices using either charge \cite{gor} 
or spin \cite{benito,loss,an} like qubits, or even both simultaneously \cite{jo,yang}. These quantum systems are of great interest because of their easy integration with existing electronics and scalability advantage \cite{ita,urda}.
Moreover,  in \cite{sanz,sza}, the quantum dynamics and
the entanglement of two electrons inside the coupled double quantum
dots were addressed, while in \cite{fan,qin,borge,sou}  aspects related
to the quantum correlations and the decoherence were investigated.  
Furthermore,  several other properties have been investigated: 
quantum teleportation based on the double quantum dots \cite{choo}, 
the quantum noise due to phonons induce steady-state in a 
double quantum dot charge qubit \cite{gia}, multielectron quantum 
dots \cite{rao} and thermal quantum correlations in two coupled 
double semiconductor charge qubits \cite{moi} were also reported.  
More recently,  a conceptual design of quantum heat machines 
has been developed using two coupled double quantum-dot 
systems as a working substance \cite{moi-1}.

In recent years, the spin-orbit interaction (SOI) in quantum dots has attracted 
much attention both theoretically and experimentally due to its potential roles 
in the quantum coherent  manipulation of a spin qubit and spintronics \cite{fron,hen}. There are two different types of SOI in a semiconductor material, i.e., The Rashba SOI due to structural inversion asymmetry \cite{ras} and Dresselhauss SOI due 
to the bulk inversion asymmetry \cite{dress}.

Interest in the SOI process has been increased in recent past years 
as a set of potential applications of the SOI process was recently reported.  
For example, the spin-orbit-coupled quantum memory of a double 
quantum dot was investigated in \cite{chot}. 
Recently, Yi-Chao Li et al. reported the influence of Rashba coupling 
in qubit gates with simultaneous transport in double quantum dots \cite{yili}, and the transport of the spin shuttling between neighboring QDs is affected by the spin-orbit interaction \cite{ginzel}.

On the other hand, quantum coherence arising from quantum superposition 
is a fundamental feature of quantum mechanics, and it has been 
widely recognized as the essence of bipartite and multipartite 
quantum correlations.
The framework for quantifying  coherence is based on taking into account  
an incoherent  basis and defining an incoherent state as one which is 
diagonal on that basis.  Several measurements have been proposed,
and their properties have been investigated in detail over the years(see
\cite{baum,Hu,stre2}, for instance). More recently, a new measure called
correlated coherence \cite{tan,tri} has been introduced to investigate
the relationship between quantum coherence and quantum correlations.
Quantum correlated coherence is a measure of coherence with removed
local parts; that is,  all system coherence  is stored entirely in quantum correlations.

Fast reliable spin manipulation in quantum dots is one of the most important challenges in spintronics and semiconductor-based 
quantum information. However, in real systems and with potential application to quantum information processing, it is crucial to understand
 the thermal  robustness of quantum correlations at high temperatures, which is one of the main goals of this paper. 

In this work, we aim to investigate the role of thermal entanglement
and the quantum correlated coherence in a single electron spin in a 
double quantum dot in the presence of an external magnetic field. 
This electron contributes to tunneling,  coupling the QDs and  spin-flip 
tunneling caused by a Rashba spin-orbit coupling. 
We assume  that the system is isolated from its respective 
electronic reservoirs, which remain  in the strong Coulomb blockade 
regime, where one electron is permitted in a double quantum dot. 
We obtained analytical solutions, which allowed us to explore in detail the 
concurrence at zero temperature  as well as the performance of the 
thermal entanglement; it is also possible to study the thermal evolution 
of the populations and thermal fidelity of the model. 
We also derived an analytical expression for the quantum correlated coherence and the difference between concurrence and quantum correlated coherence are investigated.  In addition, it is compared the thermal entanglement with a quantum correlated coherence.
Last but not least, the framework provided by the correlated coherence allows us to retrieve the same concepts of quantum discord and quantum
entanglement, providing a unified view of these correlations, where
the quantum discord is a measure of the quantum correlations going
beyond entanglement \cite{ollivier,wer}. Note that, for a multipartite
system, if the coherence of the global state is a resource that cannot
be increased, the cost of creating discord can be expressed in terms
of coherence \cite{yue,ma}.
In this paper, we study these quantifiers in a thermal bath. The processing of quantum information can be done by controlling the temperature and the Rashba effect parameter present in the double quantum dot.

The outline of this paper is as follows.  Section II defines the
physical model and the method to treat it.  Section III, briefly describes
the definition of the concurrence ($\mathcal{C}$) and the correlated
coherence ($\mathcal{C}_{cc}$). Thus the analytical expressions for them
are found. In Section IV, we discuss some of the most interesting results
like entanglement, populations, and correlated 
coherence taking into the account, the temperature effects, Rashba 
coupling and the tunneling parameter.
Finally, in Section V, we present our conclusions. 

%%%%%%%%%%%%%%%%%%%%%%%%%%%%%

\section{The model}
\begin{figure}
\includegraphics[scale=0.8]{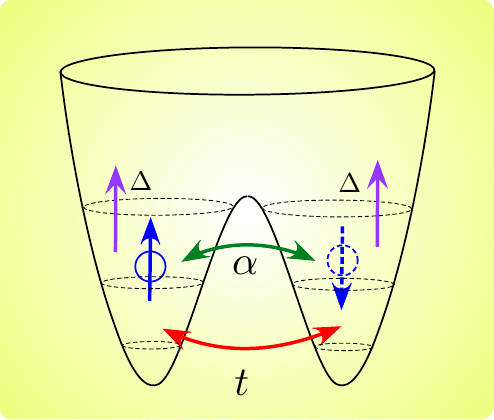}\caption{\label{fig:model} Schematic representation of the double quantum dot,
the physical model includes the Rashba interaction $\alpha$. The spin of an electron is represented by the small sphere delocalized between two quantum dots.}
\end{figure}

The setup under investigation, depicted in Fig. \ref{fig:model}, is a silicon device that consists of a double quantum dot is  filled with a single electron and has two charge configurations, with the electron located either on the left ($L$) or right ($R$) dot,  corresponding to position states labeled by $\left|L\right>$ and  $\left|R\right>$ respectively. The Hamiltonian of the double quantum dot \cite{yili} is given by
%%%%%%%%%%%%%%%%%%%%%%%%%%%%%
\begin{equation}
\begin{array}{ccc}
H & = & \frac{\Delta}{2}(\mathbb{I}\otimes\sigma_{z})+t(\tau_{x}\otimes\mathbb{I})-\alpha(\tau_{y}\otimes\sigma_{x}),\end{array}\label{eq:1}
\end{equation}
where $\tau_{x,y}$ are the Pauli matrices in the $\{\left|L\right>, \left|R\right>\}$ basis and $\sigma_{x,z}$ are the Pauli matrices describing the single electronic spin states $\{\left|0\right>, \left|1\right>\}$, $\mathbb{I}$ is the $2 \times 2$ identity matrix. Here  $\Delta$ is the Zeeman
splitting generated by a constant external magnetic field along the $z$-axis, $t$ is the strength of the tunneling coupling between
the two quantum dots, while the $\alpha$ is the spin-flip tunnel coupling due to the Rashba  SOI \cite{ras} contribution.
%%%%%%%%%%%%%%%%%%%%%%%%%%
%%%%%%%%%%%%%%%%%%%%%%%%%%%%%

The four eigenvectors of Hamiltonian (\ref{eq:1}) in the natural basis $\left\{ \left|L0\right>,\left|L1\right>,\left|R0\right>,\left|R1\right>\right\} $
are

\begin{eqnarray}
\left|\varphi_{1}\right> & = & A_{+}\left[ia_{+}\left(\left|L0\right>+\left|R0\right>\right)-\left|L1\right>+\left|R1\right>\right],\nonumber \\
\left|\varphi_{2}\right> & = & A_{-}\left[ia_{-}\left(\left|L0\right>+\left|R0\right>\right)-\left|L1\right>+\left|R1\right>\right],\nonumber \\
\left|\varphi_{3}\right> & = & B_{+}\left[ib_{+}\left(\left|L0\right>-\left|R0\right>\right)+\left|L1\right>+\left|R1\right>\right],\nonumber \\
\left|\varphi_{4}\right> & = & B_{-}\left[ib_{-}\left(\left|L0\right>-\left|R0\right>\right)+\left|L1\right>+\left|R1\right>\right],
\end{eqnarray}
where $A_{\pm}=\frac{1}{\sqrt{2}\sqrt{a_{\pm}^{2}+1}}$, $a_{\pm}=\frac{\Omega_{+}\pm\sqrt{\Omega_{+}^{2}+4\alpha^{2}}}{2\alpha}$,
$B_{\pm}=\frac{1}{\sqrt{2}\sqrt{b_{\pm}^{2}+1}}$, $b_{\pm}=\frac{\Omega_{-}\pm\sqrt{\Omega_{-}^{2}+4\alpha^{2}}}{2\alpha}$,
$\Omega_{\pm}=\Delta\pm2t$ and the corresponding eigenvalues are

\begin{eqnarray}
\varepsilon_{1,2} & = & \pm\frac{1}{2}\sqrt{\Omega_{+}^{2}+4\alpha^{2}},\\
\varepsilon_{3,4} & = & \pm\frac{1}{2}\sqrt{\Omega_{-}^{2}+4\alpha^{2}}.
\end{eqnarray}

The system state in the thermal equilibrium is described by $\rho(T)=\frac{\exp(-\beta H)}{Z}$,
where $\beta=1/k_{B}T$, with $k_{B}$ being the Boltzmann's constant,
$T$ is the absolute temperature and the partition function of the
system is defined by $Z=Tr\left[\exp(-\beta H)\right]$. 

%%%%%%%%%%%%%%%%%%%%%%%%%%%%%%%%%%

\subsection{The density operator}

%%%%%%%%%%%%%%%%%%%%%%%%%%%%%%%%%%
At thermal equilibrium,  the double quantum dot density operator $\rho$ is described as 
\begin{equation}
\rho_{AB}(T)=\left[\begin{array}{cccc}
\rho_{11} & \rho_{12} & \rho_{13} & \rho_{14}\\
\rho_{12}^{*} & \rho_{22} & \rho_{14} & \rho_{24}\\
\rho_{13} & \rho_{14}^{*} & \rho_{11} & -\rho_{12}\\
\rho_{14}^{*} & \rho_{24} & -\rho_{12}^{*} & \rho_{22}
\end{array}\right].\label{eq:5}
\end{equation}
The elements of this density matrix, after a cumbersome algebraic
manipulation, are given by

\[
\begin{array}{ccl}
\rho_{11} & = & \frac{A_{+}^{2}a_{+}^{2}e^{-\beta\varepsilon_{1}}+A_{-}^{2}a_{-}^{2}e^{-\beta\varepsilon_{2}}+B_{+}^{2}b_{+}^{2}e^{-\beta\varepsilon_{3}}+B_{-}^{2}b_{-}^{2}e^{-\beta\varepsilon_{4}}}{Z},\\
\rho_{12} & = & \frac{i[-A_{+}^{2}a_{+}e^{-\beta\varepsilon_{1}}-A_{-}^{2}a_{-}e^{-\beta\varepsilon_{2}}+B_{+}^{2}b_{+}e^{-\beta\varepsilon_{3}}+B_{-}^{2}b_{-}e^{-\beta\varepsilon_{4}}]}{Z},\\
\rho_{13} & = & \frac{A_{+}^{2}a_{+}^{2}e^{-\beta\varepsilon_{1}}+A_{-}^{2}a_{-}^{2}e^{-\beta\varepsilon_{2}}-B_{+}^{2}b_{+}^{2}e^{-\beta\varepsilon_{3}}-B_{-}^{2}b_{-}^{2}e^{-\beta\varepsilon_{4}}}{Z},\\
\rho_{14} & = & \frac{i[A_{+}^{2}a_{+}e^{-\beta\varepsilon_{1}}+A_{-}^{2}a_{-}e^{-\beta\varepsilon_{2}}+B_{+}^{2}b_{+}e^{-\beta\varepsilon_{3}}+B_{-}^{2}b_{-}e^{-\beta\varepsilon_{4}}]}{Z},\\
\rho_{22} & = & \frac{A_{+}^{2}e^{-\beta\varepsilon_{1}}+A_{-}^{2}e^{-\beta\varepsilon_{2}}+B_{+}^{2}e^{-\beta\varepsilon_{3}}+B_{-}^{2}e^{-\beta\varepsilon_{4}}}{Z},\\
\rho_{24} & = & \frac{-A_{+}^{2}e^{-\beta\varepsilon_{1}}-A_{-}^{2}e^{-\beta\varepsilon_{2}}+B_{+}^{2}e^{-\beta\varepsilon_{3}}+B_{-}^{2}e^{-\beta\varepsilon_{4}}}{Z},
\end{array}
\]
where $Z=\underset{i}{\sum}e^{-\beta\varepsilon_{i}}$.

Since $\rho_{AB}(T)$ represents a thermal state in equilibrium, the corresponding entanglement
is then called \textit{thermal entanglement}. In this paper, we consider a single electron spin in a double quantum dot with Rashba interaction. We found that, the charge qubit controlled by the interdot tunneling and the spin qubit driven by  the Rashba interaction are responsible for the thermal entanglement of the model.

%%%%%%%%%%%%%%%%%%%%%%%%%%%%%%%%%%%%%%%%%%%%

\section{Quantum Correlations}

%%%%%%%%%%%%%%%%%%%%%%%%%%%%%%%%%%%%%%%%
In this section we give a brief review concerning the definition and properties of the thermal entanglement and quantum coherence.

\subsection{Thermal entanglement }

In order to quantify the amount of entanglement associated with a given two-qubit state $\rho$, we consider concurrence $\mathcal{C}$ defined by Wootters \cite{wootters,woo} 
\begin{eqnarray}
\mathcal{C}={\rm {max}\left\{ 0,2max\left(\sqrt{\lambda_{i}}\right)-\sum_{i}\sqrt{\lambda_{i}}\right\} ,}
\end{eqnarray}here $\lambda_{i}\:(i=1,2,3,4)$ are the eigenvalues in descending
order of the matrix 
\begin{eqnarray}
R=\rho\left(\sigma^{y}\otimes\sigma^{y}\right)\rho^{\ast}\left(\sigma^{y}\otimes\sigma^{y}\right),
\end{eqnarray}
with $\sigma^{y}$ being the Pauli matrix. After straightforward calculations, the eigenvalues of the matrix $R$ can be expressed as

%\[
\begin{eqnarray}
\lambda_{1} & = & \Theta+G+\sqrt{\Xi_{+}\Sigma_{+}},\nonumber \\
\lambda_{2} & = & \Theta+G-\sqrt{\Xi_{+}\Sigma_{+}},\nonumber \\
\lambda_{3} & = & \Theta-G+\sqrt{\Xi_{-}\Sigma_{-}},\nonumber \\
\lambda_{4} & = & \Theta-G-\sqrt{\Xi_{-}\Sigma_{-}},
\end{eqnarray}
%\]
where 
\[
\begin{array}{ccl}
G & = & -2\rho_{14}\rho_{12}+\rho_{11}\rho_{24}-\rho_{13}\rho_{22},\\
\Theta & = & \rho_{11}\rho_{22}-\rho_{13}\rho_{24}+|\rho_{14}|^{2}+|\rho_{12}|^{2},\\
\Xi_{\pm} & = & 2\left(\rho_{12}\pm\rho_{14}\right)\left(\rho_{22}\pm\rho_{24}\right),\\
\Sigma_{\pm} & = & 2\left(\rho_{13}\mp\rho_{11}\right)\left(\rho_{14}\pm\rho_{12}\right).
\end{array}
\]
Thus, the concurrence of this system can be written as \cite{cao}
\begin{eqnarray}
\mathcal{C}={\rm {max}\left\{ 0,\mid\sqrt{\lambda_{1}}-\sqrt{\lambda_{3}}\mid-\sqrt{\lambda_{2}}-\sqrt{\lambda_{4}}\right\} ,}
\end{eqnarray}
In this case, the analytical expression for the thermal concurrence
is too large to be explicitly provided here, but it easy to recover following the above steps.

 %%%%%%%%%%%%%%%%%%%%%%%%%%%%%%%%%%%%%%%%%%%

\subsection{Correlated Coherence}

%%%%%%%%%%%%%%%%%%%%%%%%%%%%%%%%%%%%%%%%%%%
Quantum coherence is an important feature in quantum physics and 
is of practical significance in quantum information processing task.  
Quantum coherence in a bipartite system can  be contained both locally and in the correlations 
among the subsystems. The difference between the amount of coherence contained in the global state and the coherences that are purely local, is called \textit{correlated coherence},
$\mathcal{C}_{cc}$ \cite{tan}. For a bipartite quantum system, it
becomes 
\begin{equation}
\mathcal{C}_{cc}(\rho_{AB})=\mathcal{C}_{l_{1}}(\rho_{AB})-\mathcal{C}_{l_{1}}(\rho_{A})-\mathcal{C}_{l_{1}}(\rho_{B}),\label{eq:6}
\end{equation}
where $\rho_{A}=Tr_{B}(\rho_{AB})$ and $\rho_{B}=Tr_{A}(\rho_{AB})$.
Here, $A$ and $B$ stand for local subsystems.

In accordance with the set of properties that any appropriate measure
of coherence should satisfy \cite{baum}, a number of coherence measures
have been put forward.  Here we are concerned with the $l_{1}$-norm, it is a bona fide measure of coherence. The definition of the $l_{1}$-norm of coherence $\mathcal{C}_{l_{1}}$ is

%\[
\begin{equation}
\mathcal{C}_{l_{1}}(\rho)=\sum_{i\neq j}|\langle i|\rho|j\rangle|.
\end{equation}
%\]

Quantum coherence is a basis-dependent concept, but we can choose
an incoherent one for the local coherence, which will allow us to
diagonalize $\rho_{A}$ and $\rho_{B}$. From Eq.(\ref{eq:5}), the
reduced density matrix $\rho_{A}(T)$ will be given by 
\begin{equation}
\rho_{A}(T)=\left(\begin{array}{cc}
\rho_{11}+\rho_{22} & \rho_{13}+\rho_{24}\\
\rho_{13}+\rho_{24} & \rho_{11}+\rho_{22}
\end{array}\right).\label{eq:rha}
\end{equation}
In a similar way, we obtain 
\begin{equation}
\rho_{B}(T)=\left(\begin{array}{cc}
2\rho_{11} & 0\\
0 & 2\rho_{22}
\end{array}\right).\label{eq:rhb}
\end{equation}
In order to analyze the correlated coherence, we perform a unitary transformation
in the reduced density matrix $\rho_{A}(T)$.  Thus, the unitary matrix results in
\begin{equation}
U=\left(\begin{array}{cc}
\cos\theta & -e^{i\varphi}\sin\theta\\
e^{-i\varphi}\sin\theta & \cos\theta
\end{array}\right).\label{eq:7}
\end{equation}
So, let us have $\widetilde{\rho}_{A}(T)=U\,\rho_{A}(T)\,U^{\dagger}$. For  $\rho_{B}(T)$ it is not necessary to perform  any transformation,  the operator $\rho_{B}(T)$ is already incoherent.
On the other hand, the unitary transformation of the bipartite quantum
state $\rho_{AB}(T)$ is given by $\widetilde{\rho}_{AB}(T)=\widetilde{U}\,\rho_{AB}(T)\,\widetilde{U}^{\dagger}$,
where $\widetilde{U}=U\otimes\mathbb{I}$.

The unitary transformation will show the relationship between the
global coherence and the local coherence for several choices of 
 $\theta$ and $\varphi$ parameters. In particular, by setting $(\theta=\frac{\pi}{4},\varphi=0)$
in the Eq.(\ref{eq:7}), we obtain a matrix that diagonalize $\rho_{A}(T)$. This step provide us the basis set, where $A$
is locally incoherent. Thus, by inserting Eq.(\ref{eq:7})
into the Eq.(\ref{eq:6}), fixing $\theta=\frac{\pi}{4}$ and $\varphi=0$,
we obtain an explicit expression for correlated coherence, that is,
%\[
\begin{equation}
\mathcal{C}_{cc}(\rho_{AB}(T))=|\rho_{14}+\rho_{12}|+|\rho_{14}+\rho_{12}^{*}|+|\rho_{12}-\rho_{14}|+|\rho_{12}^{*}-\rho_{14}|.
\end{equation}
%\]

\subsection{Fidelity of thermal state}
%%%%%%%%%%%%%%%%%%%%%%%%%%%%%%%%%%%%
The mixed-state fidelity can be defined as \cite{jo,zhou}
\begin{equation}
F(\rho_{1},\rho_{2})=Tr\sqrt{ \rho_{2}^{1/2}\rho_{1}\rho_{2}^{1/2}}.
\end{equation}
This quantity measures the degree of distinguishability between the two quantum states $\rho_{1}$ and $\rho_{2}$. Conversely, the quantum fidelity between the input pure state and the output mixed state is defined by
\begin{equation}
F=\langle\psi|\rho\left|\psi\right>,
\end{equation}
where $\left|\psi\right>$ is the pure state and $\rho$ is the density operator state. This measurement provides the information of the overlap between the pure state $|\psi\rangle$  and the mixed state $\rho$.
In the our case, we will study the thermal fidelity between the ground state $\left|\varphi_{2}\right>$ and the state of the system at temperature $T$. After some algebra, one finds
\begin{equation}
F(T)=\frac{\left[a_{+}^{2}(\rho_{11}+\rho_{13})+(\rho_{22}-\rho_{24})+2ia_{+}(\rho_{12}-\rho_{14}\right]}{\left[a_{+}^{2}+1\right]}.
\end{equation}
Although this work is theoretical, a possible implementation of the device of a single electron in a double quantum dot with Rashba interaction is to consider the introduction of micromagnets in the device for spin-orbit interaction (SOI), see \cite{bor,holl}. 

%%%%%%%%%%%%%%%%%%%%%%%%%%%%%%%%%
\section{Results and Discussions}

%%%%%%%%%%%%%%%%%%%%%%%%%%%%%%%%%%%%
In this section, it is discussed the main results obtained in the
foregoing section. 

%%%%%%%%%%%%%%%%%%%%%%%%%%%%%%%%%%%%%%%%%%%

\subsection{Concurrence at zero temperature}

\begin{figure}
\includegraphics[scale=0.4]{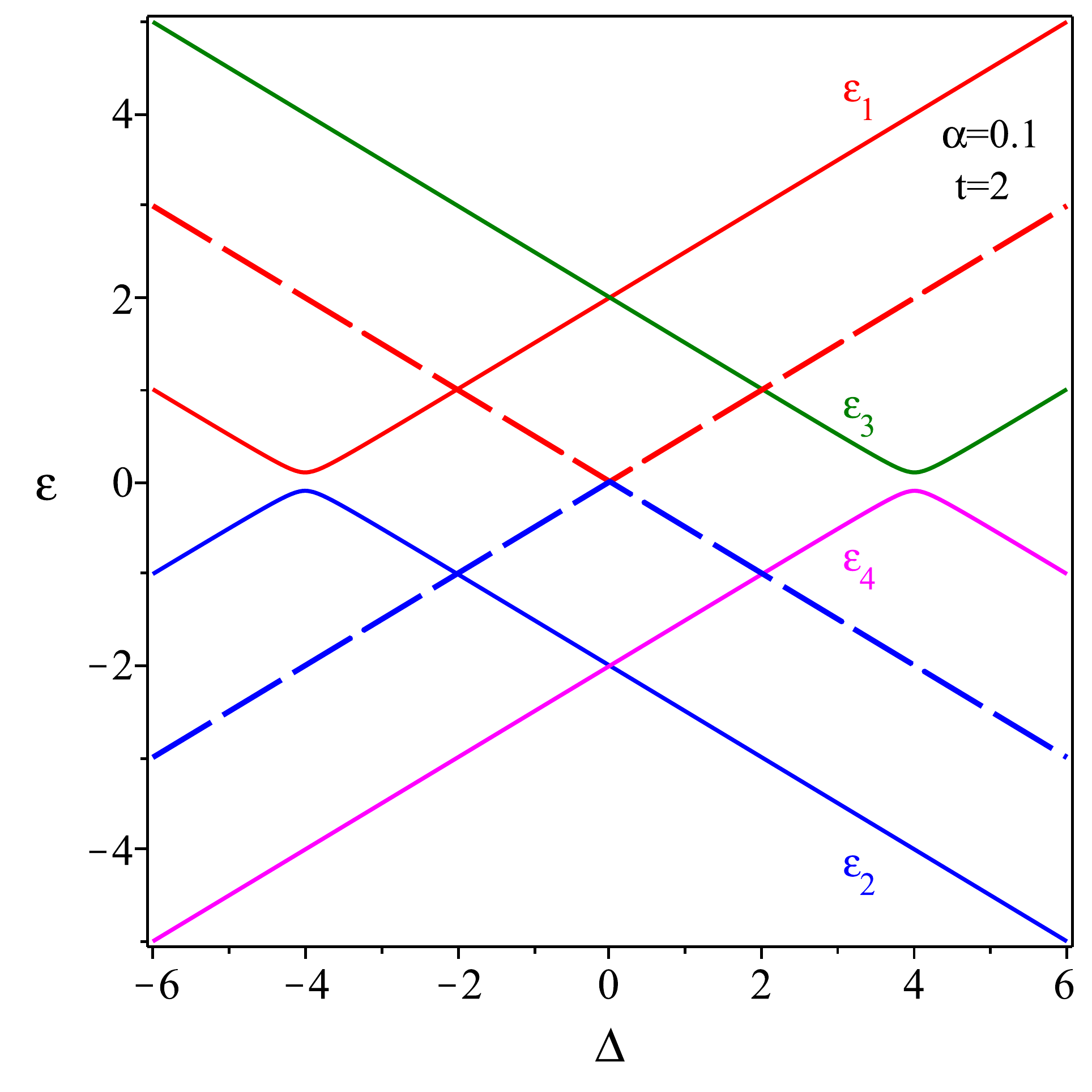}\caption{\label{fig:energy} Spectrum energy of the DQD Hamiltonian $H$ as a function of 
$\Delta$, for fixed $t=2$ and $\alpha=0.1$ (solid curves). The dashed blue line and dashed  red line show the energy levels for $t=0$ and $\alpha=0$. }
\end{figure}

Firstly, we investigate the influence of the tunneling coefficient $t$ and Rashba coupling $\alpha$ on the energy levels in zero temperature. The energy levels versus Zeeman splitting $\Delta$ is plotted in Fig. \ref{fig:energy}. Initially, we show in the same graph the two energies, each two-fold degenerate, for  $t=0$ and $\alpha=0$ as indicated by dashed lines, red ($\varepsilon_{1}=\varepsilon_{3}$) and blue ($\varepsilon_{2}=\varepsilon_{4}$), respectively.  On the other hand, for the solid curves,  the tunneling between quantum dots ($t=2$) breaks the degeneracy at $\Delta=0$. Meanwhile,  the Rashba coupling ($\alpha=0.1$) induces two anti-crossing points.  In $\Delta=4$ for energy levels $\varepsilon_{3}$, and $\varepsilon_{4}$, and in $\Delta=-4$ for energy levels $\varepsilon_{1}$ and $\varepsilon_{2}$  From  the above analysis, it is easy see that there is a strong correlation between interdot tunneling rates and degeneracy breaking of the eigenstates. As well as,  one clear signature of the spin-orbit interaction is the formation of anti-crossing points in the electron energy spectrum.  

\begin{figure}
\includegraphics[scale=0.4]{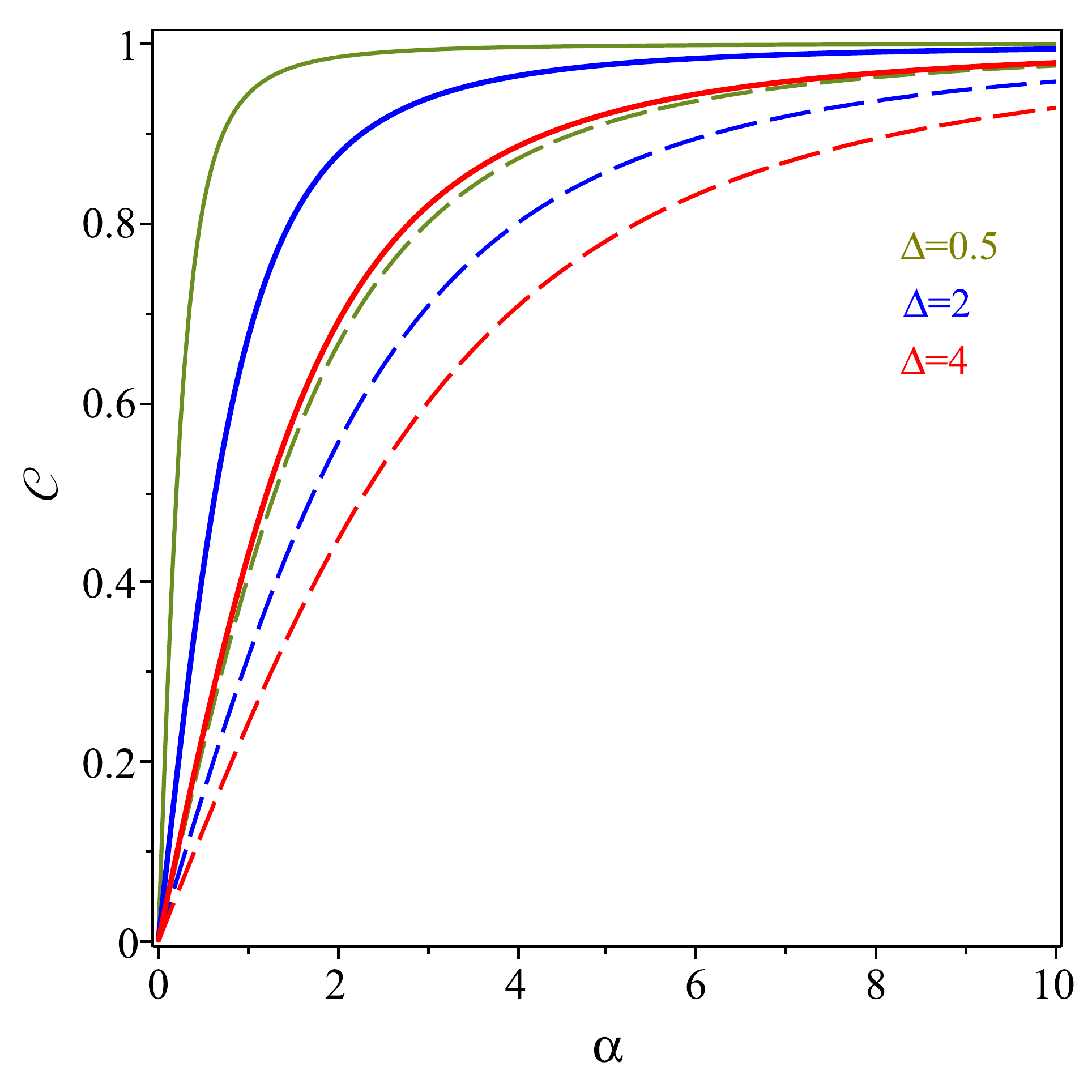}\caption{\label{fig:Con-1} The concurrence $\mathcal{C}$ as a function of 
$\alpha$, for fixed $t=0.1$ (solid curves) and $t=2$ (dashed curves) at zero temperature.  Here we choose  $\Delta=0.5$ (green curve), $\Delta=2$ (red curve) and $\Delta=4$ (blue curve).}
\end{figure}

In Fig. \ref{fig:Con-1} we plot the concurrence $\mathcal{C}$ versus Rashba coupling $\alpha$, at zero temperature for fixed $t=0.1$ (solid curves), and $t=2$ (dashed curves), assuming several values of the $\Delta$.  For tunneling parameter $t=0.1$, we observe a vigorous increase of the concurrence until reaching $\mathcal{C}\approx0.9993$ for weak Zeeman splitting $\Delta=0.5$ and
%%%%%%%%%%%%%%%%%%%%%%%%%%%%%%%%%%%%%%%%
%%%%%%%%%%%%%%%%%%%%%%%%%%%%%%%%%%%%%%%%%%%
Rashba coupling $\alpha=10$, in this case a single non-zero eigenvector that contributes to the entanglement is $\left|\varphi_{2}\right>\approx -0.491i\left(\left|L0\right>+\left|R0\right>\right)+0.508\left(-\left|L1\right>+\left|R1\right>\right)$, whereas when we consider  $\alpha \rightarrow \infty$, the ground state reduces to $\left|\varphi_{2}\right>=-0.5i\left(\left|L0\right>+\left|R0\right>\right)+0.5\left(-\left|L1\right>+\left|R1\right>\right)$ and achieving maximum concurrence  $(\mathcal{C}=1)$. Moreover,  the curves show that the entanglement between the spin-charge qubits is smaller as the Zeeman splitting increases. From the same figure, we can see that as soon as the tunneling parameter increase say $t=2$, the concurrence is weaker than for weak tunneling regime (see dashed curves).  Furthermore, still in same figure, it is observed that the concurrence is null at $\alpha=0$ for each parameter $t$ and $\Delta$ considered.  Here the unentangled ground state is given by $\left|\varphi_{2}\right>=\frac{1}{\sqrt {2}}\left(-\left|L1\right>+\left|R1\right>\right)$.

%%%%%%%%%%%%%%%%%%%%%%%%%%%%%%%
%%%%%%%%%%%%%%%%%%%%%%%%%%%%%

\subsection{Thermal Quantum Coherence}

\begin{figure}
\includegraphics[scale=0.4]{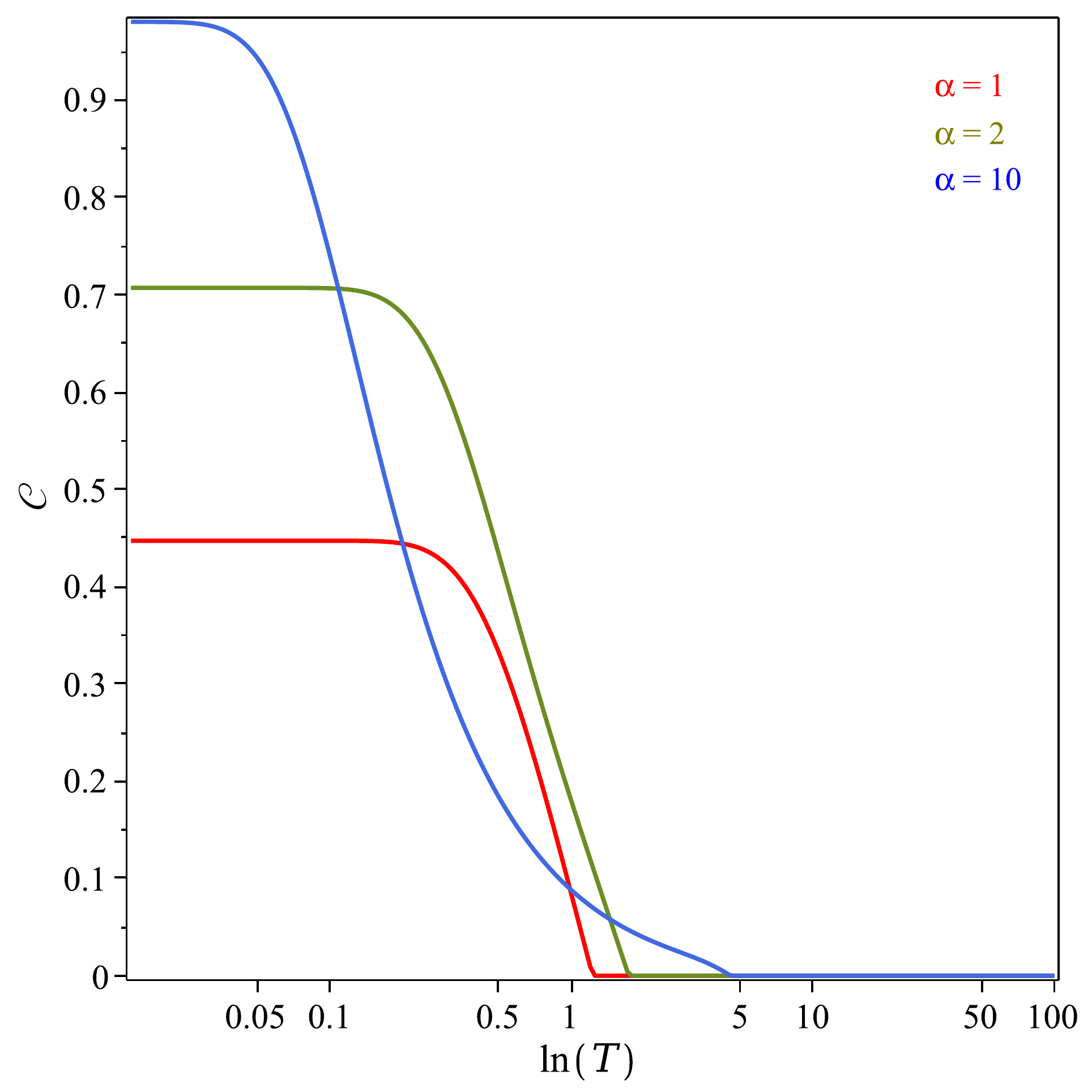}\caption{\label{fig:C-T} The concurrence $\mathcal{C}$ as a function of temperature  $T$ in the logarithmic scale, for fixed $\Delta=2$, $t=1$.  Here,  $\alpha=1$ (red
curve), $\alpha=2$ (green curve), $\alpha=10$ (blue curve).}
\end{figure}

%%%%%%%%%%%%%%%%%%%%%%%%%%%%%%%%%%%%%%%%
Firstly,  we study how the concurrence $\mathcal{C}$ is affected by temperature $T$.  In Fig. \ref{fig:C-T} we depict 
the concurrence $\mathcal{C}$ as a function of the temperature $T$ in the logarithmic scale and for different values of the Rashba coupling
$\alpha$, with $\Delta=2$ and $t=1$. It is clear to see that there are two
different regimes:  the first one corresponds to a strong Rashba coupling $\alpha=10$ (blue curve), where we can see the concurrence for $T=0$ becomes $\mathcal{C}\approx0.98$. It is also observed
that the concurrence monotonously leads to zero at the threshold temperature $T_{th}\approx4.558$. For $\alpha=2$ (green curve), the concurrence $(\mathcal{C}\approx\frac{1}{\sqrt2})$ is smaller than to the previous case at low temperature.  However, it decreases quickly as temperature raise and finally vanishes at threshold temperature $T_{th}\approx1.728$. The second one corresponds to weak Rashba coupling strength, e.g., $\alpha=1$ (red curve), 
where we obtain a weak entanglement at zero temperature $\mathcal{C}\approx0.447$, which remains almost constant at low temperature. Then,  the concurrence monotonically decreases with increasing temperature until it completely  vanishes at the threshold temperature $T_{th}\approx 1.224$. This result shows that $\alpha$ can be used for either tuning on or off the entanglement. 

\begin{figure}
\includegraphics[scale=0.45]{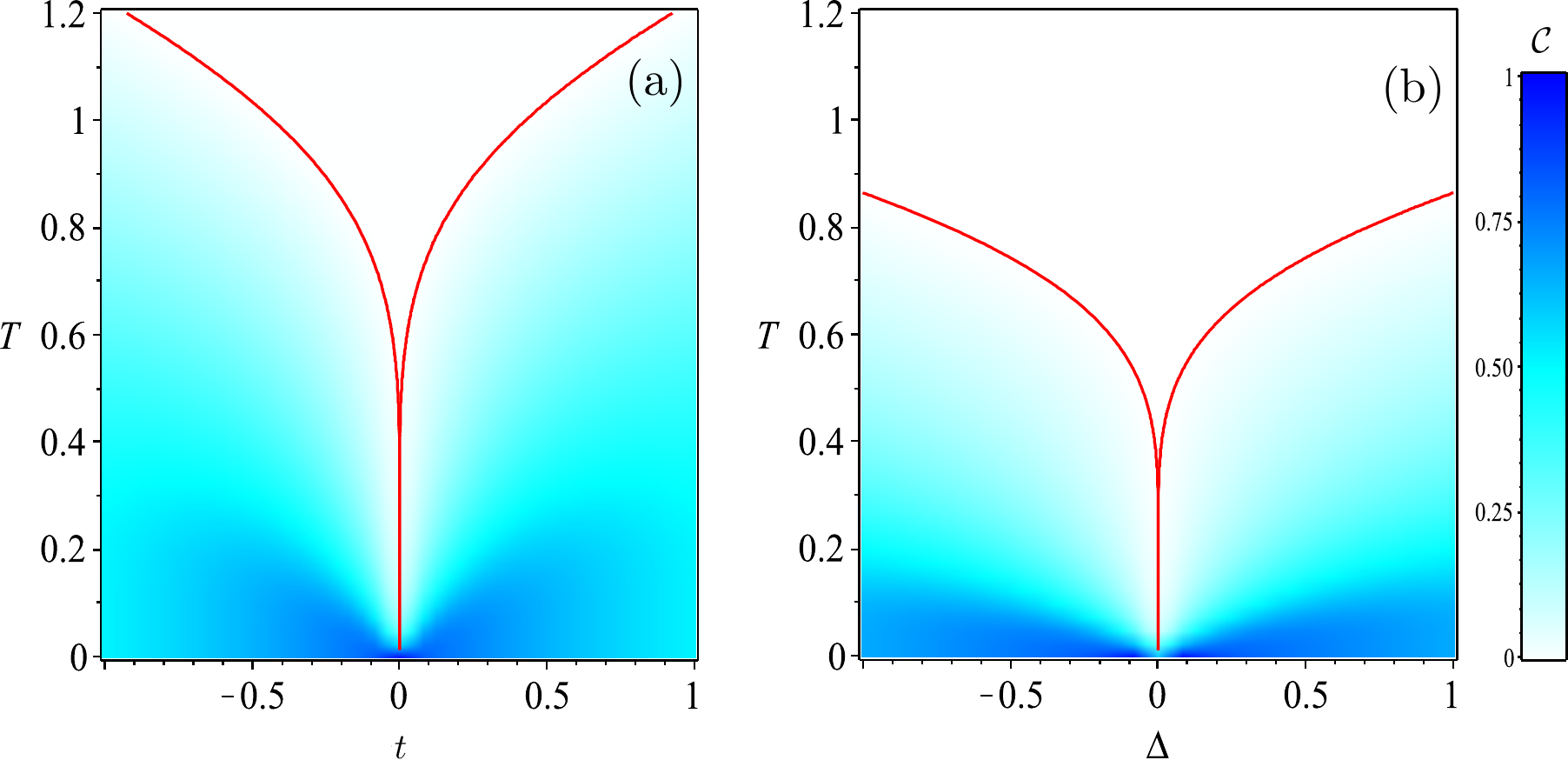}
\caption{\label{fig:density} The density plot of the thermal concurrence $\mathcal{C}$.  a) as a function of $T$ versus $t$ with $\Delta=2$ and $\alpha=1$.  b) as a function of $T$ versus $\Delta$ with $t=0.5$ and $\alpha=1$.  In these figures, red solid curve is the contour between the entangled region (blue) and the disentangled region (white). }
\end{figure}

In Fig. \ref{fig:density}(a), we illustrate the density plot of  concurrence $\mathcal{C}$ as a function of $T$ and $t$, for 
fixed values of $\Delta=2$ and $\alpha=1$. The blue color corresponds to the entangled region, while the white color 
corresponds to the unentangled region. One interesting feature observed here is that the system is strongly entangled around $t=0$ and at 
low temperatures. There is a threshold temperature above which the entanglement becomes zero.  We  also observed that the concurrence gradually decreases with the increase of the tunnel effect parameter, which indicates that the tunnel effects weakens the quantum entanglement.  
Furthermore, a similar density plot for the concurrence is reported in Fig. \ref{fig:density}(b) as a function  
 of $T$ and $\Delta$ for fixed values of $t=0.5$ and $\alpha=1$. Still, in the same panel, we can notice that when the Zeeman splitting is null,  the model is weakly entangled in a low temperature region. 
 But quickly, the concurrence disappears due to the 
 thermal fluctuations as the temperature increases. Additionally, the density plot also shows that the entanglement is strong for weak Zeeman splitting values at zero temperature, but the entanglement decreases as the Zeeman parameter increases. 
On the other hand, when $T$ increases, the concurrence $\mathcal{C}$ decreases rapidly until achieving the threshold temperature, above which the thermal entanglement
becomes null. 

%%%%%%%%%%%%%%%%%%%%%%%%%%

%%%%%%%%%%%%%%%%%%%%%%%%%%%

%%%%%%%%%%%%%%%%%%%%%%%%%%%%%%%%%%%%%%%%%%%%%%

\begin{figure}
\includegraphics[scale=0.40]{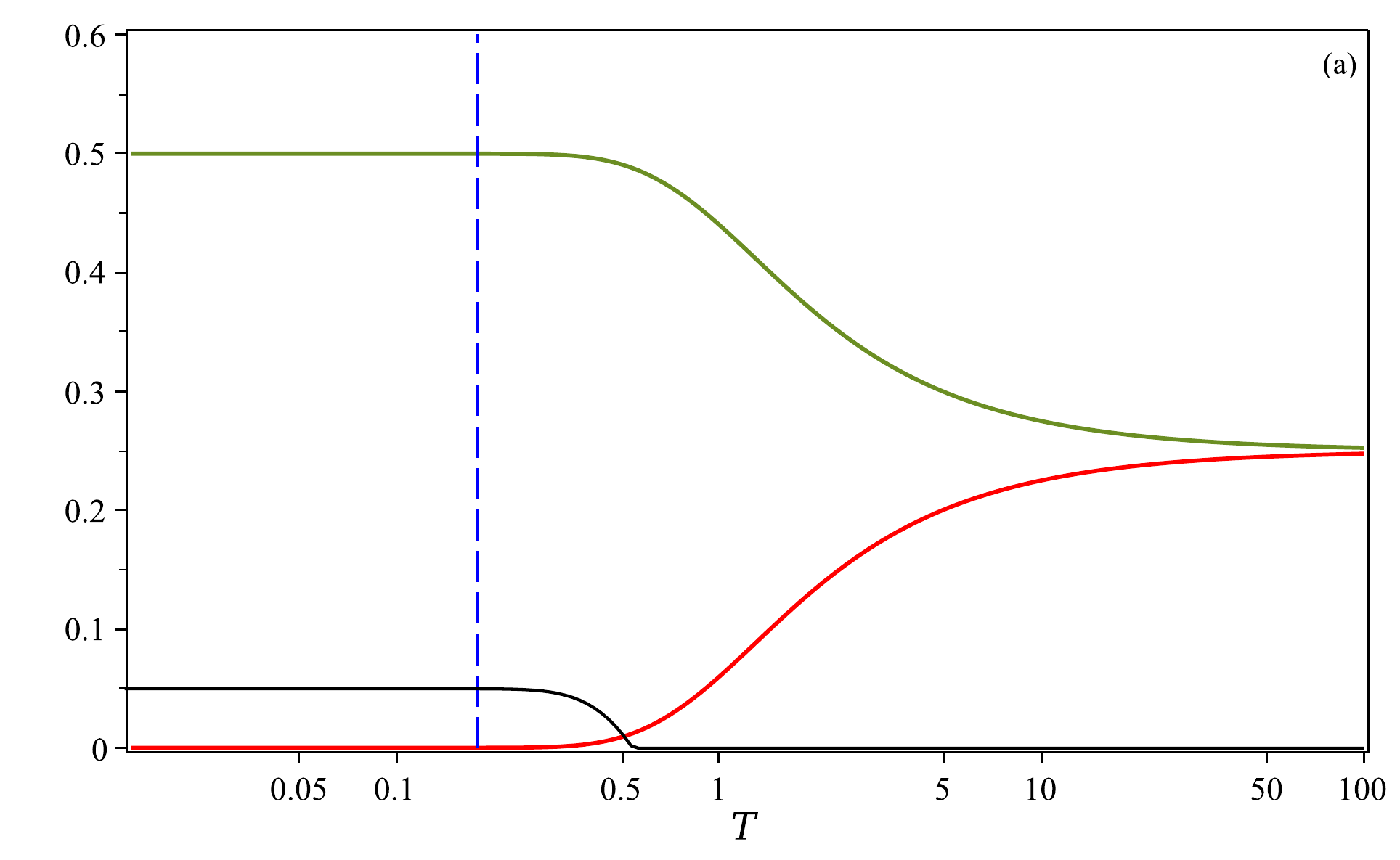}
\includegraphics[scale=0.40]{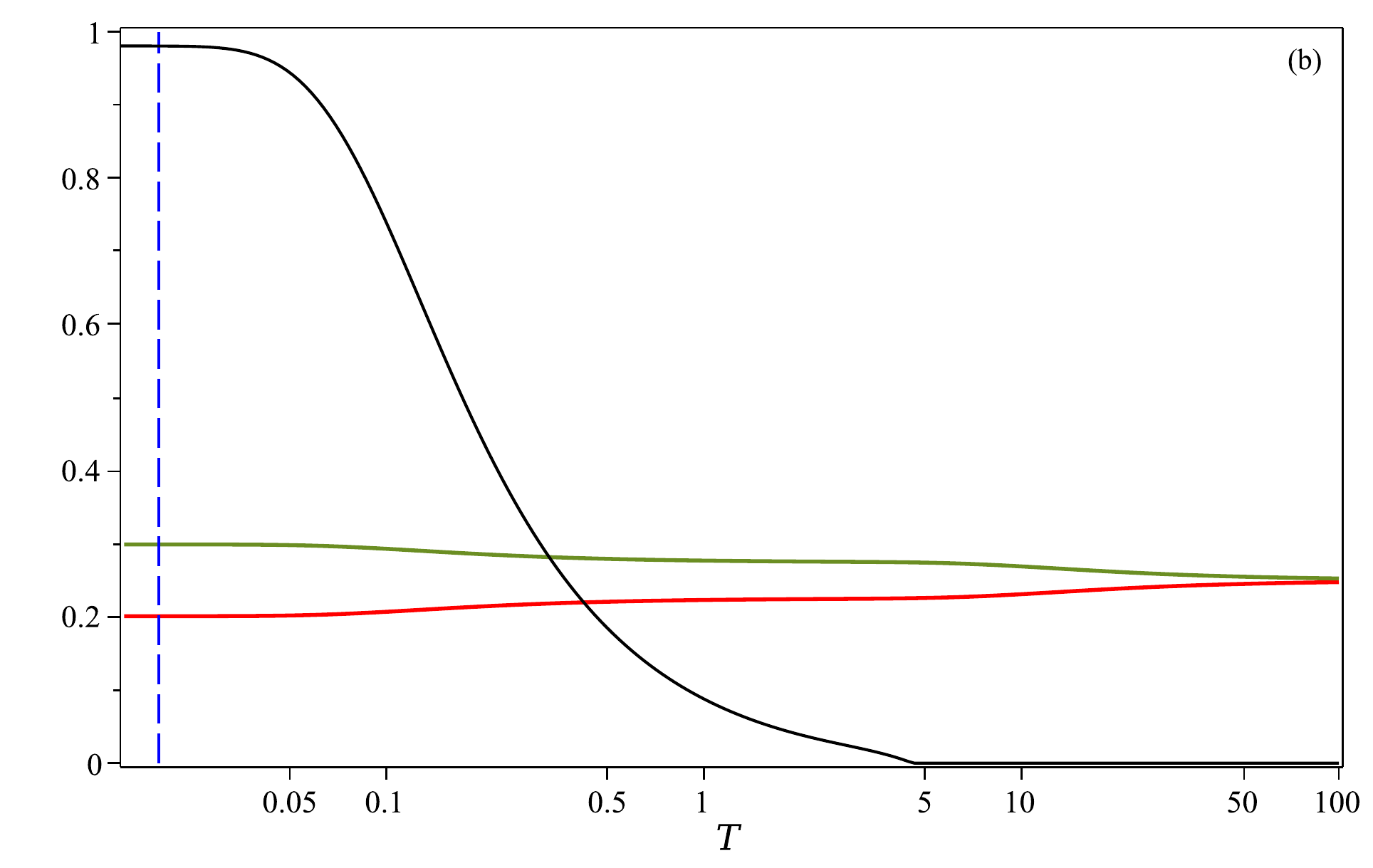}
\caption{\label{fig:popu} The thermal effects on the population $\rho$ and concurrence  $\mathcal{C}$. Here, the red curve corresponding to $\rho_{11}$ and green curve corresponding to $\rho_{22}$, while black curve represents to $\mathcal{C}$. The  parameters are set as $\Delta=2$, $t=1$. (a) $\alpha=0.1$, (b) $\alpha=10$.}
\end{figure}

In Fig. \ref{fig:popu},  the thermal effects on populations $\rho_{11}$ (red curve), $\rho_{22}$ (green curve) and concurrence (black curve),  are reported for two values of the Rashba coupling. In this figure, the blue dashed line shows the threshold temperature going from the region of constant concurrence to the region where concurrence monotonously decreases as the temperature increases, this threshold temperature also describes the beginning of population change.  In Fig. \ref{fig:popu}(a) for the Rashba coupling $\alpha=0.1$. We have observed that for low temperatures, the population and concurrence remain constant in a small range of temperature, in this region we find that the populations are $\rho_{11}\approx 0.003$ (red curve) and $\rho_{22}\approx0.4996$ (green curve). These results suggest that the weakly entangled qubits are in the ground state $\left|\varphi_{2}\right>\approx-0.017i\left(\left|L0\right>+\left|R0\right>\right)-0.7068\left(-\left|L1\right>+\left|R1\right>\right)$ for low temperature regimes, so the concurrence is $\mathcal{C}\approx0.0499$. In this figure, the blue dashed line shows the threshold temperature is $T_{th}\approx0.1777$. Thus, we found that quantum entanglement is sensitive to population change as a consequence of increasing temperature.  On the other hand, in Fig.\ref{fig:popu}(b) for a strong Rashba coupling $\alpha=10$, we observe a sudden 
increase of $\rho_{11}$ which attains the value $\rho_{11}\approx0.2$ (red curve), a decrease for $\rho_{22}\approx0.3$ (green curve) and the concurrence reaches the value $\mathcal{C}\approx0.9805$ at low temperatures, this concurrence is constant until the threshold temperature $T_{th}\approx0.0191$, see the blue dashed line. Therefore, due to thermal fluctuations, the populations undergo a change and concurrence decreases until it disappears.  In any case,  with increasing temperature regardless of the value of the Rashba coupling,  the population corresponding to the $\rho_{11}$ state 
increases, while the population $\rho_{22}$ decreases 
until at higher temperature the eigenstates are distributed equally,  reaching the value $0.25$.  

\begin{figure}
\includegraphics[scale=0.40]{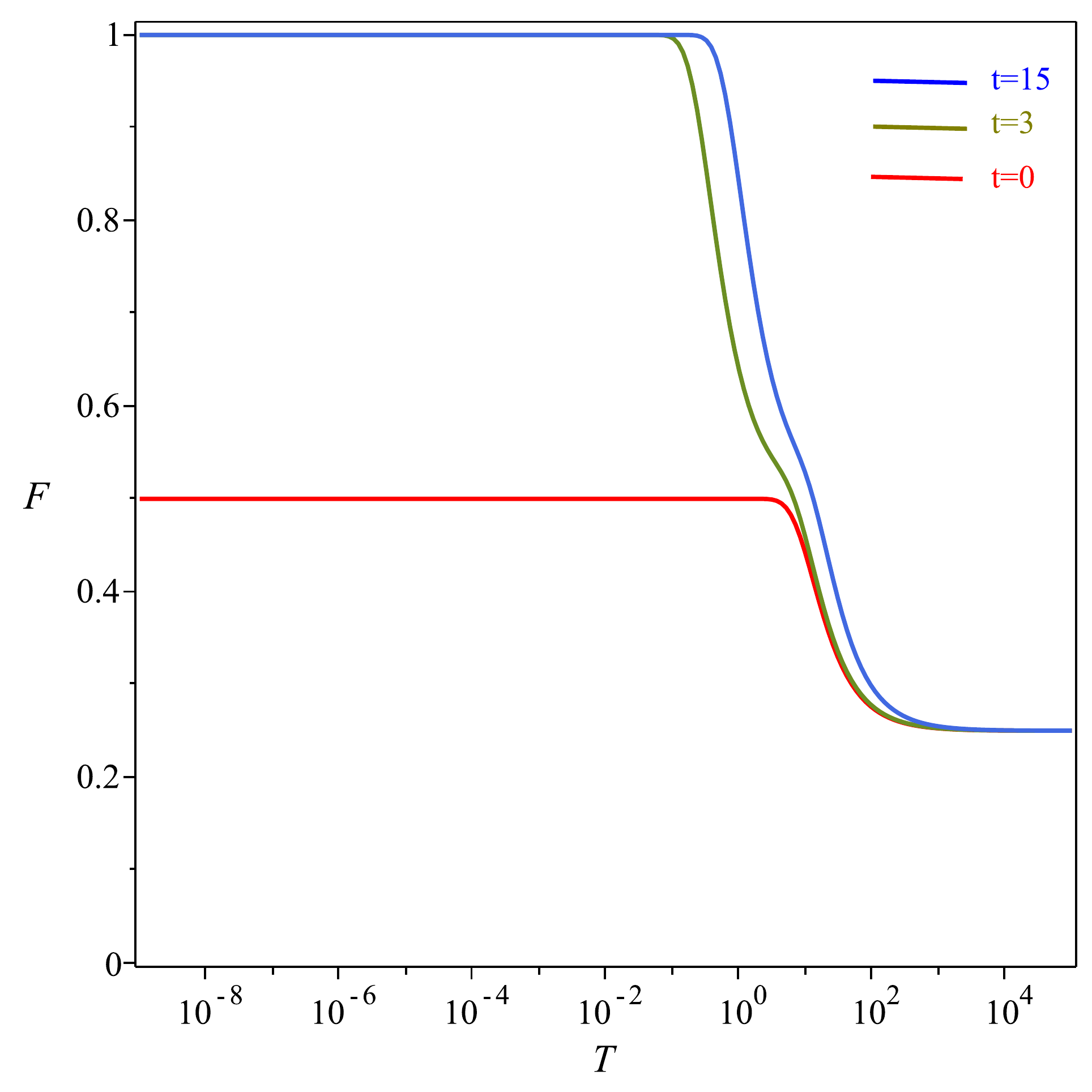}
\caption{\label{fig:7} The thermal fidelity $F$ as a function of temperature. Here, the red curve corresponding to tunneling coupling $t=0$, green curve corresponding to $t=3$, while blue curve represents the case $t=15$. The  parameters are set as $\Delta=2$, $\alpha=10.0$}
\end{figure}

In Fig.\ref{fig:7}, we plot the fidelity $F$  between the ground state $\left|\varphi_{2}\right>$  and the thermal state $\rho_{AB}(T)$ as a function of temperature $T$ in the logarithmic scale.  We can see that  the mixed-state fidelity approaches ground-state fidelity i.e. $F=1$, when the temperature leads to zero. On the other hand, when the temperature increases the  ground state mixes with the excited states, allowing the fidelity to decrease monotonically as the temperature increases. It is also observed that for $T=0$, the figure exhibits the change of the fidelity $F=0.5$ (red curve), since the ground states become the degenerate states $\left|\varphi_{2}\right>$ and $\left|\varphi_{4}\right>$ for fixed tunneling parameter $t=0$.

%%%%%%%%%%%%%%%%%%%%%%%%%%%%%%%%%%%%%%%%%%%%
%%%%%%%%%%%%%%%%%%%%%%%%%%%%%%%%%%%%%%%%%%

%%%%%%%%%%%%%%
%%%%%%%%%%%%%%%%%%%%%%%%%%%%%%%%%%%%%%%%

\begin{figure}
\includegraphics[scale=0.37]{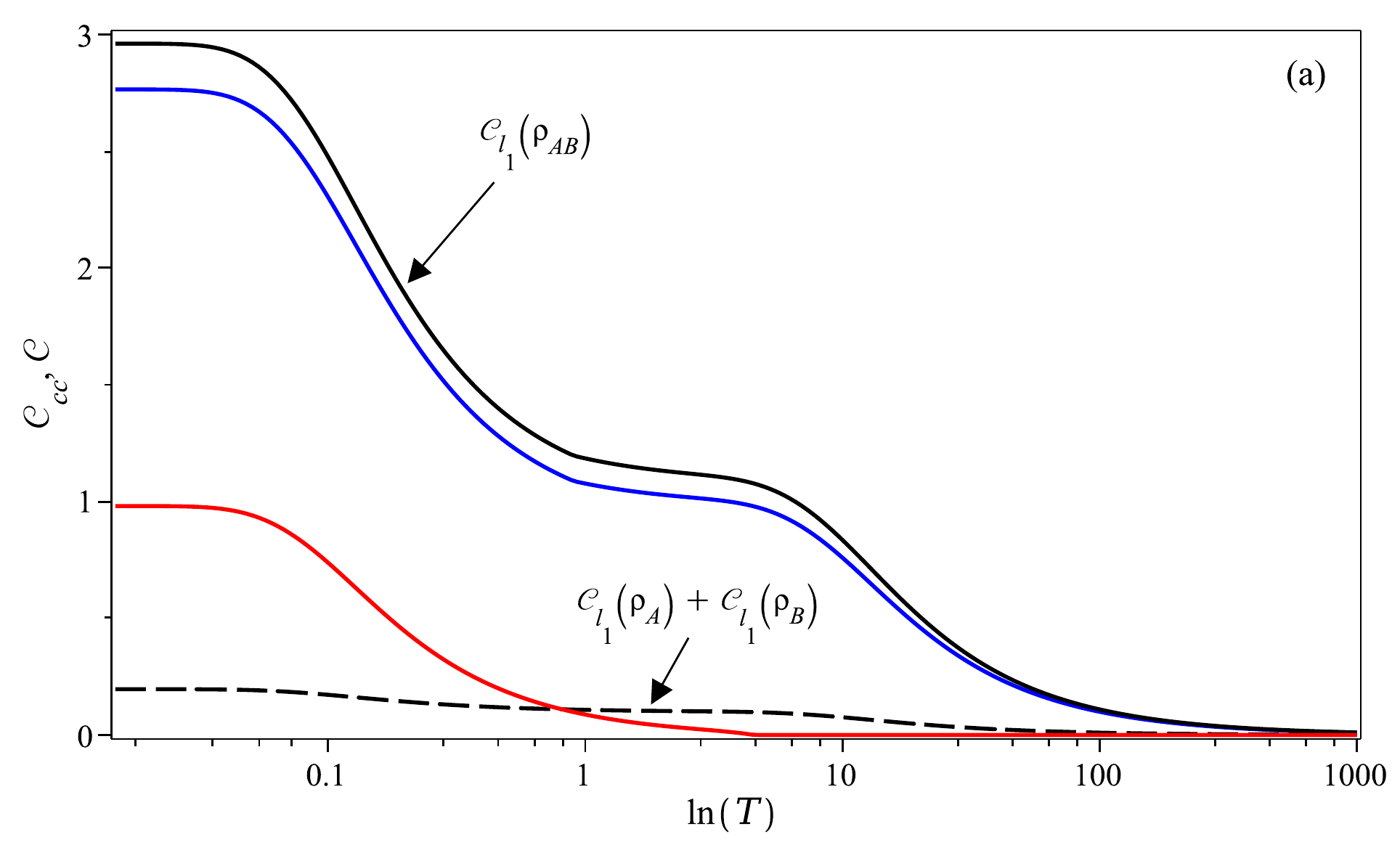}
\includegraphics[scale=0.37]{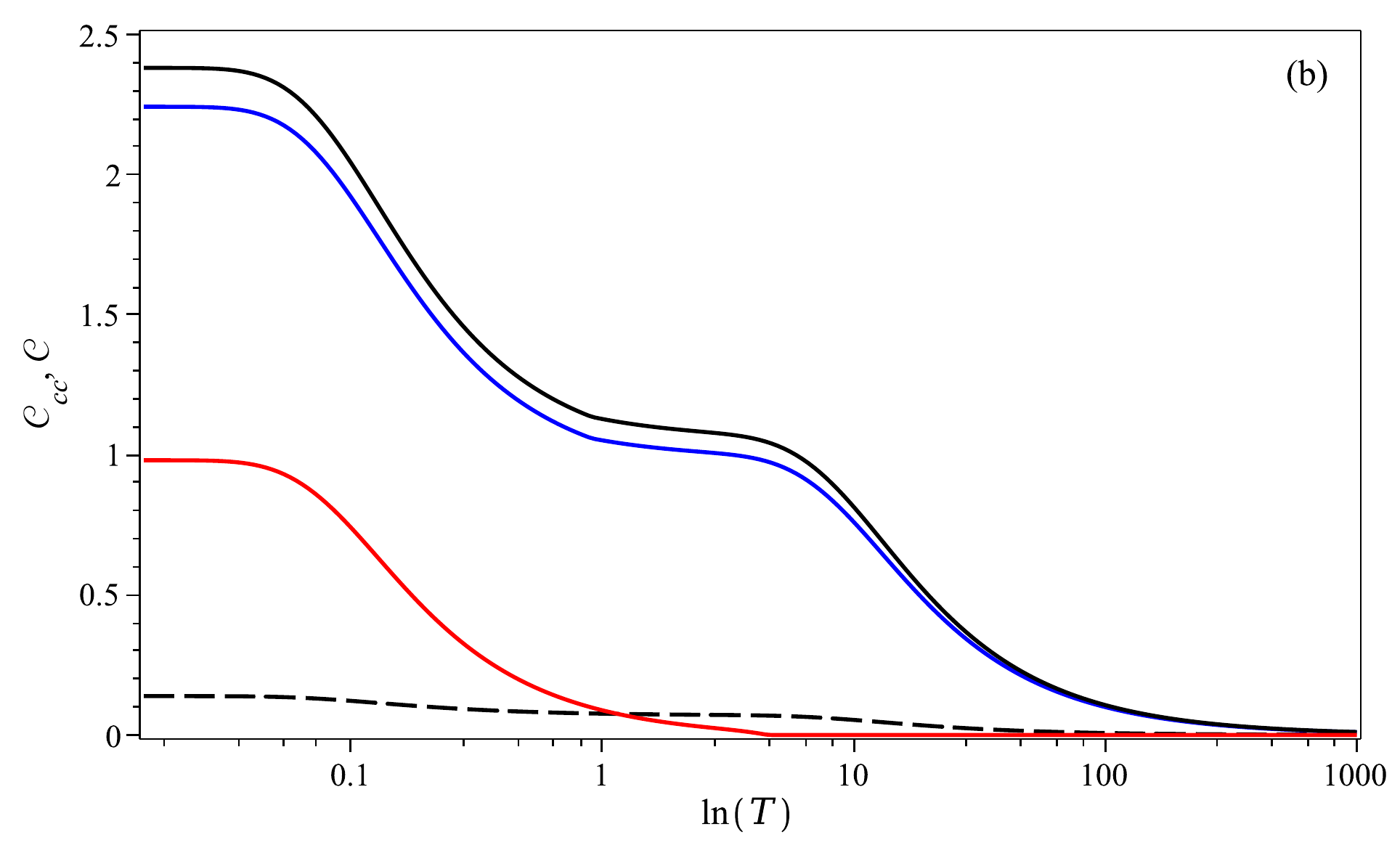}
\includegraphics[scale=0.37]{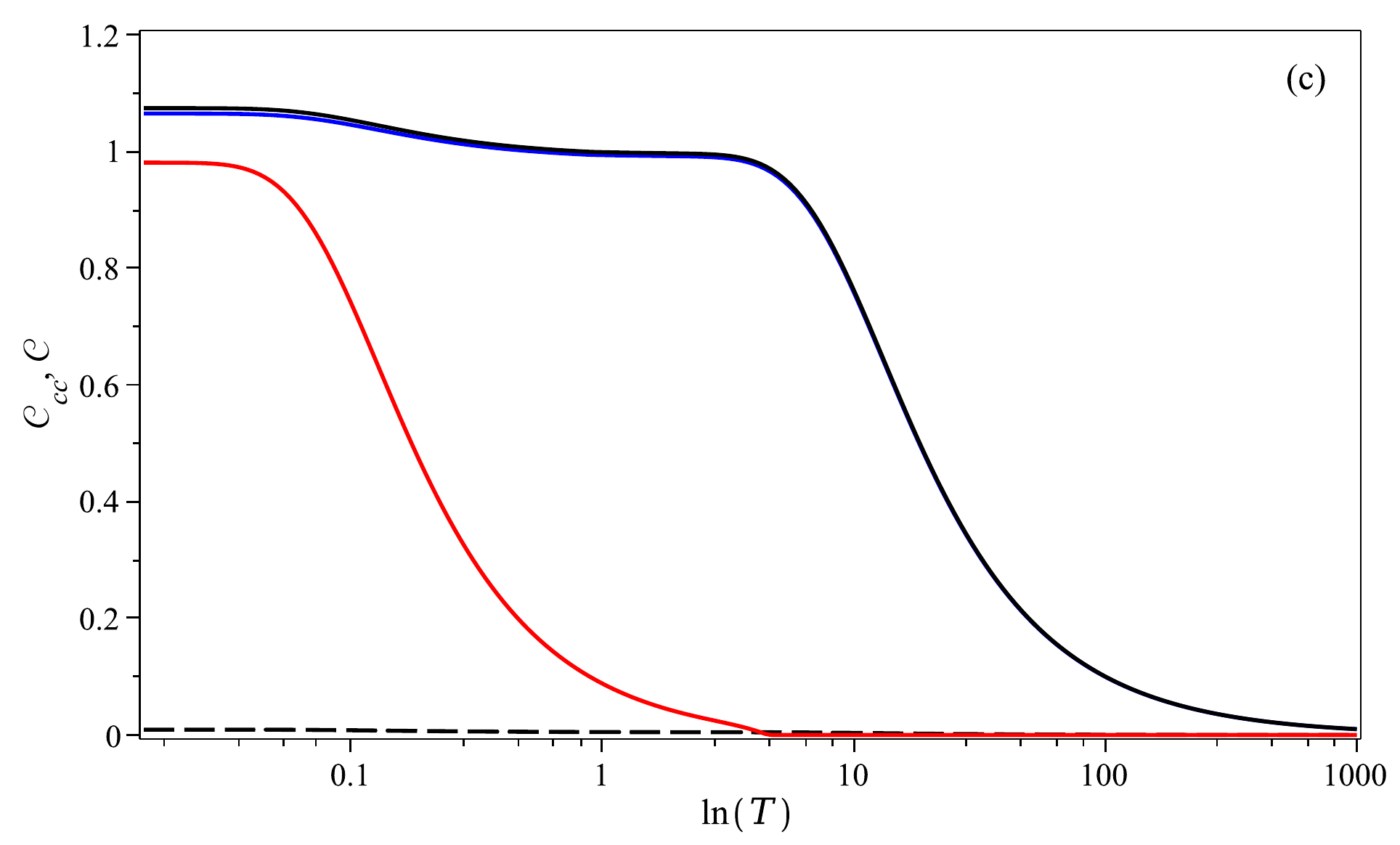}
\includegraphics[scale=0.37]{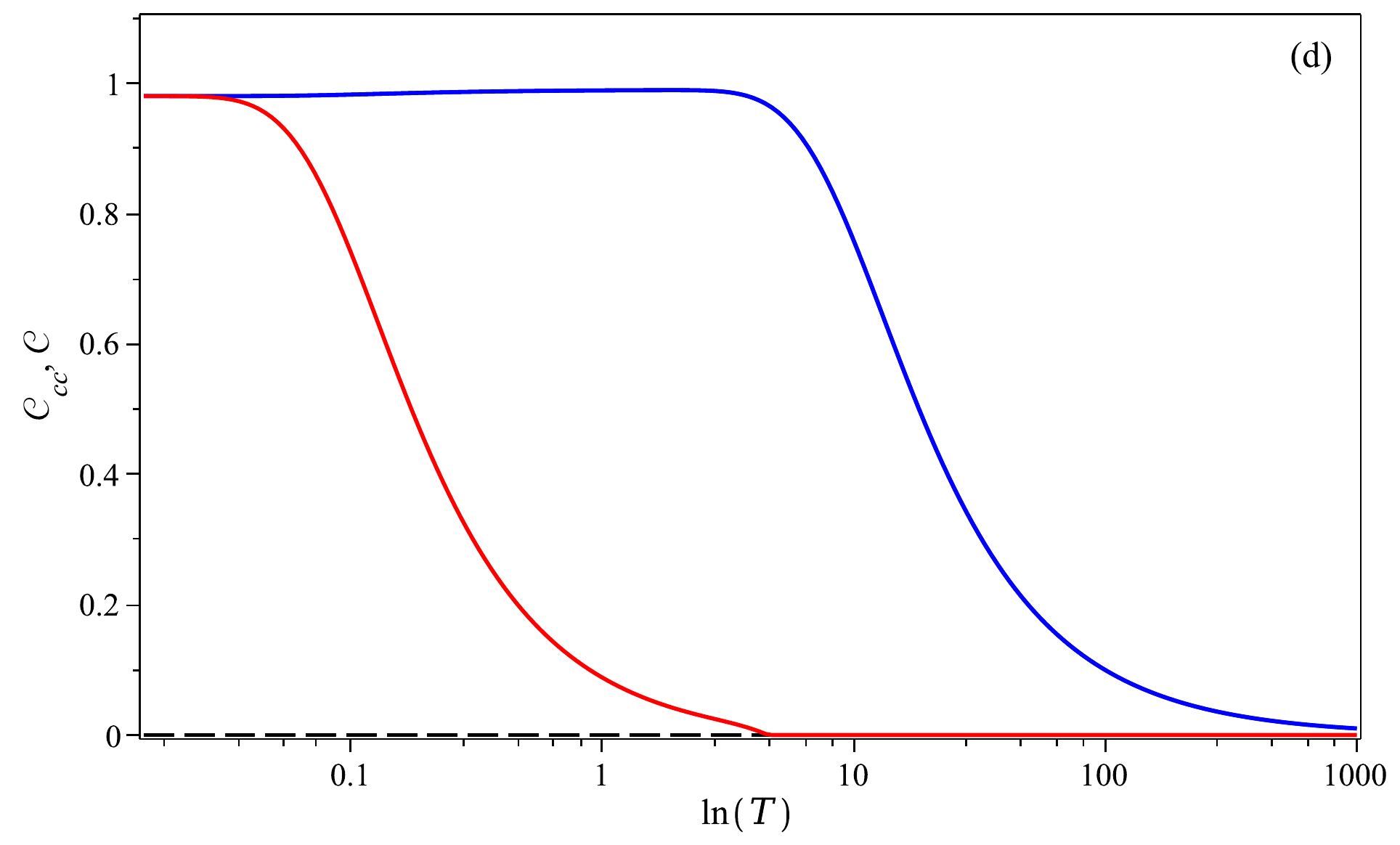}

\caption{\label{fig:8} Correlated coherence $\mathcal{C}_{cc}$ (blue solid curve) and concurrence $\mathcal{C}$ (red solid curve)  versus $T$ 
in the logarithmic scale for different values of $\theta$. In particular we set 
$\Delta=2$,  $t=1$, $\alpha=10$ and $\varphi=0$. (a) $\theta=0$, (b) $\theta=(\frac{\pi}{8})$, (c) $\theta=0.95(\pi/4)$, (d) $\theta=(\pi/4)$.}
\end{figure}

Finally, in Fig. \ref{fig:8}, we give the plot of correlated coherence and the concurrence as a  function of temperature at a fixed value of the tunneling parameter $t=1$, Rashba coupling  $\alpha=10$, Zeeman parameter $\Delta=2,$ and for different values of the parameter $\theta$. Note that in these figures, we include the curves of total quantum coherence $\mathcal{C}_{l_{1}}(\rho_{AB})$ (black curve) and the local quantum coherence $\mathcal{C}_{l_{1}}(\rho_{A})+\mathcal{C}_{l_{1}}(\rho_{B})$ (black dashed curve) for a better understanding of these amounts.  
In Fig. \ref{fig:8}(a), we plot the correlated coherence and the concurrence as a function of temperature
$T$, in the basis of the eigenenergies which corresponds to the angle
$\theta=0$ and to $\varphi=0$ in the transformation $U$(see Eq.
\ref{eq:7}). These curves show that, for $T\rightarrow0$, the correlated
coherence $\mathcal{C}_{cc}$ (solid blue curve) is higher than the
thermal entanglement $\mathcal{C}$(solid red curve). The difference
between them is the untangled quantum correlation (quantum discord). We can also notice the presence of a plateau in the correlated coherence in this low temperature regime, this is due to the fact that the correlated coherence of the ground state $(\left|\varphi_{2}\right> )$ is weak affected by thermal fluctuations in this regime.
From this figure, it is also easy to see that, as the temperature increases, the entanglement
(red curve) decays up to threshold temperature $T_{th}\approx4.5$, while the total quantum coherence gradually decreases as the temperature
increases. In Fig. \ref{fig:8}(b), we repeat the analysis for a starting angle of $\theta=\frac{\pi}{8}$. Here, we observed a decrease 
in local quantum coherence  that accompanies the lowering of total quantum coherence, which follows as a consequence of the reduction of correlated coherence. Interestingly,  the behavior of correlated coherence, as well as total and local quantum coherence qualitatively follows the same pattern as in Fig. \ref{fig:8}(a).
In Fig. \ref{fig:8}(c), we choose $\theta$ close to
$\frac{\pi}{4}$, $\left(\theta=0.95\frac{\pi}{4}\right)$ and $\varphi=0$, for this choice of the $\theta$ parameters, we observed a dramatic decrease in correlated coherence. In addition, we can see that the local quantum coherence (dashed
black curve) is almost null.  Then, it can be seen that the correlated coherence almost  entirely constitute the total quantum coherence 
(solid black curve)  for this particular choice of $\theta$. On the other hand, for high temperatures 
and after the concurrence and the local coherence have disappeared, 
the total quantum coherence is composed solely of non-entangled quantum correlations.

%%%%%%%%%%%%%%
%%%%%%%%%%%%%%%%%%%%%%%%%%%%%%%%%%%%%%%%

To recover the independence of the correlated coherence basis, we choose the local natural basis of $\rho_{A}$, which is obtained by choosing $\theta=\frac{\pi}{4}$ and $\varphi=0$ (the reduced density matrix $\rho_{B}$ is already diagonal). Thus,  in Fig.  \ref{fig:8}(d), the concurrence and 
quantum coherence are analyzed for the incoherent basis 
$\theta=\frac{\pi}{4}$ and $\varphi=0$.
It is interesting to note that at low temperatures, the entangled quantum correlations 
of the system are stored entirely in the quantum coherence; this indicates 
that,  in this case, the correlated coherence captures all the thermal entanglement information. As the temperature increases, 
the thermal fluctuations generate 
a slight increase in quantum coherence, while the entanglement
decays and disappears at the threshold temperature, $T\approx4.5$. Finally, the correlated coherence leads monotonically to zero.

%%%%%%%%%%%%%%%%%%%%%%%%%%%%%%%%%%%%%%%%%%%%
%%%%%%%%%%%%%%%%%%%%%%%%%%%%%%%%%%%%%%%%

\section{Conclusions}

%%%%%%%%%%%%%%%%%%%%%%%%%%%%%%%%%%%
This paper considers a device composed of a single electron in a double quantum dot subjected to a homogeneous magnetic field and a spin-flip tunnel coupling due to the Rashba spin-orbit interaction in a thermal bath. The proposed model was exactly solved and the effects of temperature on  quantum coherence were analyzed.  Firstly,  the spectrum energy is discussed. It is shown that the tunneling parameter contributes to breaking the energy degeneracy, while the Rashba coupling induces anti-crossing phenomena in the electron energy spectrum.  In this model, we have investigated the thermal entanglement and correlated coherence.  We show that thermal entanglement for a single electron is possible via charge and spin qubits in a silicon double quantum dot.  Furthermore, our results suggest that the Rashba parameter turns on the thermal entanglement and be tuned conveniently.   We also have investigated the influence of the Rashba coupling on the population and concurrence. These results show that are sensitive to temperature and Rashba coupling, particularly, in the regime of low temperatures, the concurrence and populations form plateaus. However, with increasing temperature, the populations undergo changes in their behavior, while the concurrence decreases, this is a consequence of thermal fluctuations. Additionally,  we present an analysis of the thermal fidelity between the fundamental state and the thermal states, and we showed that the fidelity is maximum for low temperatures, while,  with increasing temperature, the fidelity decreases monotonically due to the mixture between the ground state and the excited states. Moreover,  we found a direct connection between entanglement
and quantum coherence.  We ultimately compare the concurrence with 
correlated coherence, which is responsible for quantum correlations. Quantum coherence is a base-dependent concept.  We have choosen an incoherent basis for the local coherence ($\theta=\frac{\pi}{4}$, $\varphi=0$), obtaining the correlated coherence. In particular, we reported that the correlated 
coherence measure is equal to the concurrence for low temperatures.
The thermal entanglement must then be viewed as a particular case
of  quantum coherence.  Furthermore, the model showed a peculiar 
thermally-induced increase of correlated coherence due to the 
emergence of non-entangled quantum correlations as the 
entanglement decreased. When $T$ is high enough, the quantum entanglement disappears  as thermal fluctuation dominates the system. 
Overall, our results highlight that the Rashba coupling can be used successfully to enhance the thermal performance of quantum entanglement.
Then, we can safely conclude that quantum coherence is more robust than entanglement under the effect of a thermal bath. The results also suggest that correlated coherence may potentially be a more accessible quantum resource in comparison to entanglement, and that this is something worth investigating in future work.

 %%%%%%%%%%%%%%%%%%%%%%%%%

\section{Acknowledgments}

%%%%%%%%%%%%%%%%%%%%%%%%%
This work was partially supported by CNPq, CAPES and Fapemig. Moises Rojas
would like to thank National Council for Scientific and Technological  Development (CNPq) - Grant No. 317324/2021-7.

\end{document}